 \documentclass[12pt,preprint]{aastex}

\usepackage {epsf}
\newcommand \m {M$_\odot$}

\newcommand\kms{km~s$^{-1}$}
\newcommand\degree{$^{\rm o}$}

\begin{document}

\title{The Shape of Cas A}

\author{J. Craig Wheeler, Justyn R. Maund, Sean M. Couch}
\affil{Department of Astronomy, University of Texas, Austin, TX 78712}
\email{wheel@astro.as.utexas.edu,jrm@astro.as.utexas.edu,smc@astro.as.utexas.edu,}

\begin{abstract}

Based on optical, IR and X-ray studies of Cas A, we propose a geometry 
for the remnant based on a ``jet-induced" scenario with significant 
systematic departures from axial symmetry. In this model, the main jet 
axis is oriented in the direction of strong blue-shifted motion at an 
angle of 110 - 120\degree\ East of North and about 40 - 50\degree\ 
to the East of the line of sight. Normal to this axis would be an 
expanding torus as predicted by jet-induced models. In the proposed 
geometry, iron-peak elements in the main jet-like flow could appear 
``beyond" the portions of the remnant rich in silicon by projection 
effects, not the effect of mixing. In the context of the proposed 
geometry, the displacement of the compact object from the kinematic 
center of the remnant at a position angle of $\sim$169\degree\ can be 
accommodated if the motion of the compact object is near to, but slightly 
off from, the direction of the main ``jet" axis by of order 30\degree. 
In this model, the classical NE ``jet," the SW ``counter-jet" and 
other protrusions, particularly the ``hole" in the North, are 
non-asymmetric flows approximately in the equatorial plane, e.g., out 
through the perimeter of the expanding torus, rather than being 
associated with the main jet. We explore the spoke-like flow in the 
equatorial plane in terms of Rayleigh-Taylor, Richtmyer-Meshkov and 
Kelvin-Helmholz instabilities and illustrate these instabilities with 
a jet-induced simulation.  

\end{abstract}


\section{Introduction}

The star we now recognize as the supernova remnant Cassiopeia A
probably exploded in 1680 \citep{thors01,fesen07}. 
It has been an astrophysical mystery ever since its re-discovery 
by radio astronomers \citep{ryle48} and has been the target of 
increasingly sophisticated studies over a range of wavelengths 
in the radio \citep{and91,keo96,liszt99}, optical 
\citep{fesen96,fesen01,fesenetal01,fesen06b,fesen06c,morse04}, 
IR \citep{krause05,ennis06}, and X-ray \citep{markert83,hughes00,willingale02,
willingale03,hwang00,hwang01,hwang04,delaney04,laming06}.   
The projected image of Cas A shows a prominent jet-like structure
in the NE and a ``counter-jet" diametrically opposed in the SW. 
Other protrusions and interruptions in the images are notable in 
images at the various wavelengths. The first image taken with 
the {\it Chandra Observatory} dramatically revealed the long-sought 
compact object. This dim object has yet to be given a firm physical 
identification as a neutron star or black hole. The compact 
object is displaced from the kinematic center of the remnant 
\citep{fesen06b} to the South, nearly perpendicular to the locus 
of the ``jet/counter-jet" structure. X-ray studies of the remnant 
have led to the discovery that iron-rich regions lie beyond 
silicon-rich regions \citep{hughes00}, suggesting some 
sort of inversion of the expected ``onion-skin" structure of 
the progenitor star. Fine-scale turbulence might mix the 
originally stratified composition and lead to Fe at larger radii
than Si, but it is not clear that such a process can lead
to a large scale segregation of the compositions, as the
observations suggest. 
 
\citet{laming06} consider the possibility that the NE/SW ``jets" 
represent a spherical explosion into an inhomogeneous CSM with 
``cavities" so that expansion is faster into those cavities, as 
suggested by \citet{blondin96}. They find that such a model cannot 
give a sufficiently high density of plasma at high enough temperature. 
The ejecta expand so rapidly that the density is too low either 
for appreciable electron-ion equilibration to raise the electron 
temperature or to sufficiently ionize the plasma as observed for 
the NE jet.  Laming et al. consider a jet-induced model based on 
the calculations of \citet{kho99} and find that such a model can 
provide the required conditions of ionization age and electron 
temperature for the collisional ionization equilibrium knots at 
the jet tip and the non-ionization equilibrium further back in the 
jet stream. Laming et al. argue that there should be a substantial 
amount of cold plasma at the ``jet" head that has been cooled by 
radiative and adiabatic losses. This material could be composed of
Fe, but is not necessarily so.  Laming et al. note that the blast 
wave is not seen in the direction of the NE ``jet." The NE jet is 
thus not overdense compared to the stellar envelope, as it was for 
the jet-induced supernova models of \citet{kho99} \citep[see also][]
{hof01,kho01}. Laming et al. find that the kinetic energy in the NE 
``jet" is a relatively modest $10^{50}$ ergs, perhaps insufficient 
to represent the major power source that caused the supernova 
explosion.  This leaves open the basic physical mechanism of the explosion. 

\citet{hwang03} argue for an ejecta mass of about 2 \m\ and a total 
mass of the star at the time of the explosion of about 3 \m.
\citet{chev03} have made the case for a clumpy wind that may
account for the quasi-stationary flocculi and that the explosion 
might have corresponded to a Type IIn or Type IIb supernova, in 
current typology.  \citet{young06} argue that the progenitor was a 
star of 15 - 20 \m\ on the main sequence that lost its hydrogen 
envelope to a binary companion and exploded with a mass of 4 - 6 \m.

Among the elements that must be explained to fully account for the
shape of Cas A are:
\begin{itemize}
\item A basic filametary morphology (Fig. 1a)
\item Quasi-stationary flocculi \citep{fesen01}
\item Fast-moving knots of oxygen and their concentration at
certain angles \citep{fesen06c}
\item The appearance of ``rings" at various parts of the perimeter
(Fig. 1b)
\item The asymmetric distribution of mass \citep{willingale03}
\item The distribution of silicon and other intermediate mass 
elements \citep{hwang04}
\item The concentration of iron to the SE and NW, especially its
appearance at larger projected radii toward the SE than silicon 
and its apparent absence in the NE ``jet" \citep{hughes00,hwang04} 
\item A prominent blue-shifted region to the SE \citep{markert83,
hwang01,willingale02}
\item The prominent ``jet" and ``counter-jet" structure in the NE/SW
\citep{fesen01,hwang04}
\item The location and direction of motion of the compact object
\citep{fesen06b}
\item The lack of a prominent pulsar wind nebula around the compact 
object \citep{fesen06a}
\item The expansion velocities and energetics as a function of distance,
angle, and composition \citep{willingale02,hwang04,fesen06b,laming06}
\item The presence or absence of a companion star \citep{fesen06a}
\end{itemize}

As a step toward accounting for these properties we consider 
the physics and morphology implied by a ``jet-induced" 
supernova model. The motivation for such a model has arisen 
in the context of recent developments in the study of the 
spectropolarimetry of core-collapse supernovae 
\citep{wan03,maund07a,maund07b,maund07c,WW08} that show that
core core collapse supernovae are routinely aspherical and
frequently display a prominent axisymmetry. One obvious
way to induce such an asymmetry is to explode the supernova
with bi-polar jets \citep{kho99}.
\citet{WMA07} point out that while bi-polar ``jet-like" flow
is common in core collapse supernovae, evidence for non-axisymmetric 
structure is also prominent. The presence of non-axisymmetric
morphology is manifested as ``loops" in the plane of the Stokes 
parameters Q and U.  This non-axisymmetric structure was 
prominent in SN~1987A \citep{crop88} and is now recognized to
be composition-dependent, with different species ejected in 
different directions.  Deeper study of Cas A may help to inform 
our interpretation of the spectropolarimetric data and vice versa. 
A basic picture of a bi-polar explosion with significant non-axisymmetric
flow may apply in general to core-collapse explosions. Here we
explore whether such a picture can help to illuminate the 
morphology of Cas A.

\section{Proposed Geometry}

As an organizing principle, we will adopt here the ansatz that 
the explosion of Cas A was jet-induced without going into detail as 
to the physical origin of the jet(s).  A basic jet-induced model has 
certain generic features: 1) principal high-velocity flow along an
axis, presumably the rotational axis of the progenitor star
and of the new-born compact remnant; 2) bow shocks generated by
the jets, and a convergence of those bow shocks and associated
flow onto the equator with 3) the subsequent expulsion of the bulk 
of the stellar core and mantle in a toroidal configuration \citep{kho99,
hof01,kho01}. For simple models in which the ``up" and ``down"
jets are identical, the resulting fundamental geometry comprises
a jet and torus structure. Here we will explore what orientation
of this model, and what departures from its basic depiction,
may account for the observed structure, kinematics, and composition
distribution of Cas A. A key component of the jet-induced models
is the distribution of the original ``onion-skin" layers of the
massive star progenitor \citep{hof01,kho01,maeda03}. One expects 
that the jet will be predominantly composed of iron-peak matter 
arising from deep within the progenitor. The bulk of the star, 
composed of the outer layers of Si, Ca, O, He, and perhaps some 
H, that surrounded the iron core will be compressed and expelled 
in the expanding equatorial torus. 

If a jet/torus geometry pertains to Cas A, a principal question to 
ask is the direction of the main jet flow. \citet{laming06}
have established that the the NE ``jet" does represent a 
jet-like flow, but its rather feeble energy suggests that this
may not be the principal axis of the explosion. We note that in
the jet-induced model, the energy originally injected in a 
relatively narrow solid angle in the jets is redistributed
throughout the stellar envelope. The amount of energy remaining
in the original jet direction is model dependendent. Thus how 
measurements of energy and velocities in a given direction
constrain models is not obvious.  \citet{markert83} presented 
{\it Einstein} data that showed a substantial blue shift in 
the SE direction as projected on the sky, not along the classic 
NE ``jet." \citet{hwang01} and \citet{willingale02}  presented the 
same feature in {\it Chandra} and {\it XMM Newton} X-ray Doppler shift maps. 
\citet{willingale03} identify a torus with an axis oriented 
approximately in this direction.  \citet{dewey06} present another 
version of a Cas A X-ray Doppler map coupled with transverse proper 
motions from \citet{delaney04} to yield a 3D map of a selection of 
X-ray emitting knots. This map (comprising only 17 points, but 
promising the richness to come) shows again that, as viewed from the 
top, normal to the line of sight, the major blue-shifted component 
is oriented at about 45\degree\ to the East (clockwise) from the 
line of sight to the observer, with a red shifted component at about 
225\degree. From the observer's point of view, the knots with highest 
blue shift form a rough ``ring" with a rather large opening angle 
of $\sim$ 30\degree. This again suggests a major flow pattern at a 
projected position angle (counterclockwise from North on the
plane of the sky) of about 125\degree\ as illustrated in Fig. 1a. 
This direction is also marked by a concentration
of Fe that extends beyond the Si-rich material in that direction
\citep{hughes00,hwang04}. Here we assume that the direction indicated 
by these observations represents the projected main axis of the jet 
in the jet/torus geometry. We suggest that the main jet axis is 
then oriented to the SE at a position angle of about 125\degree,
roughly 40 - 50\degree\ to the East and 20 - 30\degree\ to the South 
with respect to the line of sight. 

This geometry suggests an alternative explanation for
the observation of iron-peak matter beyond the silicon-rich
material \citep{hughes00}. Rather than a literal mixing
of the matter through Rayleigh-Taylor or other processes,
the apparent distribution of matter of different composition
could be affected by projection effects. In particular, 
in the proposed geometry, the main jet would be iron rich.  
The equatorial flow would tend to be Si rich (also O and Ca 
rich). In this configuration, the iron in the jet in the SE
direction approaching the observer would tend to be faster, 
but it would also appear to be ``outside" the Si in the SE
portion of the slower moving torus because of projection 
effects. This might give a natural explanation for the 
apparent ``overturn" of the Fe compared to the Si, without 
requiring ``overturn," {\it per se}. 

Because of entrainment effects there could be some high-velocity 
Si in the axial jet flow and some Fe in the equatorial flow. 
Note also that the initial energy in the jet will be dissipated 
throughout the mantle and envelope and that rapid flow in the 
direction of the jet could lead to adiabatic expansion and 
cooling so that a significant portion of the jet flow would be
difficult to detect in current X-ray observations, as found for 
the NE ``jet" by \citet{laming06}. 

\section{The Compact Object}

\citet{fesen06b} have shown that the position of the compact 
object in Cas A is located to the South at a position angle 
of $169 \pm 8.4$\degree\ from the kinematic center of the explosion
determined by tracing the proper motion of expanding optical
emission knots (see Fig. 1a). This has posed a special puzzle 
for a picture in which the classic NE jet and SW counter-jet 
represented the principal axis of the explosion. One would then
expect, on general grounds under the jet-induced ansatz, 
the motion of the compact object to be 
roughly along that axis. Instead the motion inferred by the 
displacement of the compact object from the kinematic center is 
nearly normal to the NE/SW axis.  In the geometry proposed here, 
the PA of the compact object requires some, but only modest, 
misalignment from the main axis of the jet in the SE direction. 

Consider the geometry of Cas A in spherical polar coordinates,
with the observed position angles projected onto the plane of the sky.
In spherical polar coordinates, let $\theta$ be the latitudinal angle,
measured from East (counterclockwise) of North and $\phi$ be the 
longitudinal angle measured in the plane of the observer, such that 
$\cos(\phi)=1$ is toward the observer and $\cos(\phi)=-1$ is in 
the opposite direction with $\phi$ increasing to the East (clockwise
as viewed from the North).  With ${\bf \widehat{\underline{i}}},
 {\bf \widehat{\underline{j}}}, {\bf \widehat{\underline{k}}}$
unit vectors corresponding to the observer direction ($\phi$ = 
0\degree), East ($\phi$ = 90\degree), and North, respectively, 
the position vectors of the jet (subscript {\it J}) and compact 
object (subscript {\it C}) are then:
\begin{equation}
\widehat{{\bf r}}_{J}=\sin\theta_{J}\cos\phi_{J}{\bf
\widehat{\underline{i}}}+\sin\theta_{J}\sin\phi_{J}{\bf
\widehat{\underline{j}}}+\cos\theta_{J}{\bf \widehat{\underline{k}}},
\end{equation}
\begin{equation}
\widehat{{\bf r}}_{C}=\sin\theta_{C}\cos\phi_{C}{\bf
\widehat{\underline{i}}}+\sin\theta_{C}\sin\phi_{C}{\bf
\widehat{\underline{j}}}+\cos\theta_{C}{\bf \widehat{\underline{k}}}.
\end{equation}
The angle, $\chi$, between the jet and compact object is given by:
\begin{equation}
\hat{{\bf r}}_{J} \cdot \hat{{\bf r}}_{C} = |\hat{{\bf r}}_{J}| |\hat{{\bf
r}}_{C}| \cos \chi,
\end{equation}
such that
\begin{equation}
\label{angle1}
\cos \chi = \sin \theta_{J} \cos \phi_{J} \sin \theta_{C} \cos \phi_{C}
+\sin \theta_{J} \sin \phi_{J} \sin \theta_{C} \sin \phi_{C} + \cos
\theta_{J} \cos \theta_{C}.
\end{equation}
The observed position angle on the sky, $\theta_{p,C}$, measured 
East of North (counterclockwise), is related to the latitudinal 
angle, $\theta_{C}$, of the 3D model of Cas A by:
\begin{equation}
\label{constraint}
\tan \theta_{p,C} = \tan \theta_{C} \sin \phi_{C}.
\end{equation}
Equation \ref{constraint} can be solved to give $\sin \phi_{C}$
and $\cos \phi_{C}$ in terms of $\sin \theta_{C}$ and $\cos \theta_{C}$.
When substituted into Eqn. \ref{angle1}, this gives $\cos \chi$
as a function of $\cos \theta_{C}$: 
\begin{equation}
\label{angle}
\cos \chi = \sin \theta_{J} \cos \phi_{J} 
   \left[1 - \left(1 + \tan^2 \theta_{p,C}\right) \cos^2 \theta_{C} 
   \right]^{1/2} + \left(\sin \theta_{J} \sin \phi_{J} \tan \theta_{p,C}
   + \cos \theta_{J} \right) \cos \theta_{C}.   
\end{equation}
Figure \ref{chi} gives $\chi$ as a function of the unknown angle,
$\theta_C$, for a range of plausible values of $\theta_J$ and $\phi_J$
for the nominal value of the position angle of the compact object,
$\theta_{p,C}$. Note that $\theta_{C}$ is bounded on one side by 
motion essentially on the line of sight corresponding to 
$\theta_{C}$ = 90\degree\, for which $\phi_{C}$ = 0 and $\cos \chi 
= \sin \theta_{J} \cos \phi_{J}$.  On the other extreme the motion 
at the observed position angle could be in the plane of the sky 
corresponding to $\phi_{C}$ = 90\degree\ and $\theta_{C} = 
\theta_{p,C}$, for which $\cos \chi = \sin \theta_{J} 
\sin \phi_{J} \sin \theta_{p,C} + \cos \theta_{J} \cos \theta_{p,C}$.
While there are ranges of parameter space with large values of 
$\chi$, there are clearly ranges that give a modest value, $\chi 
\lesssim 30$\degree. Fig. \ref{chi} shows that for a given choice 
of the jet angles, $\chi$ has a minimum at a specific value of 
$\cos \theta_{C}$.  Figure \ref{chimin} gives a plot of $\chi_{min}$ 
versus $\theta_{C}$ for $\theta_J = 120$\degree\ and a range of values 
of $\phi_J$ and $\theta_{p,C}$.

Figures \ref{chi} and \ref{chimin} show that while there are valid
portions of parameter space for which the separation angle would
be large, there is an ample and reasonable range of parameter space
for which the direction of motion would be near to that of
the proposed main jet. In particular, if the main axis of 
the jet is about 40 - 50\degree\ East and 20 - 30\degree\ South 
of the line of sight ($\theta_{J} = 110 - 120$\degree), then the 
motion of the compact object would only have to be of order 
30\degree\ off the main axis of the jet to point at a projected 
position angle $\theta_{p,C} \sim 169$\degree. In this proposed 
geometry, the PA of the compact object would be nearly, if not 
exactly, aligned with the axis of the main SE jet and 
nearly orthogonal to the NE/SW ``jet" axis. While this
departure of the motion of the compact object from
the principal axis requires further explanation, it is
completely in keeping with the tilt of the rings in
SN 1987A from the apparent axis of symmetry \citep{wang02}
and with the dispersion of proper motions of young pulsars from 
the axes of symmetry of their remnants \citet{ng06}, and
perhaps with growing evidence from spectropolarimetry for
``tilted-jet" models of core collapse wherein the jets
defining bi-polar axes are not exactly aligned with the
dominant geometry of the progenitor 
\citep{maund07a,maund07b,maund07c}. 
  
\section{Jets and Holes}

What, then, are we to make of the classic NE ``jet" and SW
``counter-jet" structure? Beside these features, Cas A shows 
other evidence for ``rings" and ``holes" in optical imaging as 
spectacularly displayed in new HST ACS images as displayed in the 
Hubble Heritage collection (Fig. 1b; Fesen, private communication 2007;
http://heritage.stsci.edu/2006/30/index.html)
In particular, there is a prominent ring in the North (on the
far side from the observer) and a smaller ring within that, 
suggesting the rims of ``holes." There are other smaller, but 
distinct, ring-like features in the SE and SW. Clearly, Cas A does 
not correspond to a simple axially-symmetric, jet-induced flow.  

The outward flow and shocks involved in a supernova explosion 
are, in many circumstances, subject to Rayleigh-Taylor (RT) 
and Richtmyer-Meshkov (RM) instabilities. In addition, any 
hydrodynamic flow that is not strictly spherically 
symmetric will involve shear and the possibility of 
Kelvin-Helmholz (KH) instabilities.
All such instabilities could contribute to the breakdown
of strict axial symmetry. In the jet-induced models of
\citet{kho99} the jet was, in fact, subject to the KH
instability, but the growth time was long compared to the
propagation time out of the core, so the structure, 
although computed in full 3D, maintained the axisymmetric
structure of the initial conditions despite the presence
of the instability. Another way of breaking symmetry
is to have axial jets of different strength. \citet{kho99} 
computed two axial jets of identical nature, but discussed
the possibility that the ``up" and ``down" jets could 
have different energy and momenta, with the difference
in the momenta being delivered to the compact object
in the form of a ``kick." This imbalance in jet properties
could both break the mirror symmetry that was imposed
in the calculations of \citet{kho99} and cause a
deviation of the directions of the jets and the recoil 
of the compact object, with possible implications for 
Cas A and other core collapse supernovae.  

We thus also hypothesize that while the main jet and 
counter-jet that actually triggered the explosion of Cas A
are to the SE/NW, the classical NE/SW ``jet/counter-jet"
structure and the northern ``hole" are secondary effects
resulting from instabilities in the equatorial plane of 
relatively modest energy $\sim 10^{50}$ ergs. In particular,
we postulate that these features represent faster ``spoke" 
or ``finger" ``jet-like" flow in the plane of the expanding 
torus due to various instabilities encountered by the
expanding toroidal flow.  

In the jet-induced model, the equatorial, toroidal 
component will inevitably be subject to KH 
instabilities as it propagates out into the mantle and envelope. 
The growth time is shorter for smaller wavelengths, but 
the shortest wavelengths will be suppressed by any restoring
force as might be rendered by embedded magnetic fields. 
It is thus difficult to determine the characteristic scale
of these streaming instabilities in the absence of an
appropriate 3D numerical calculation. \citet{hunger03}
performed relevant 3D calculations with both jet-like and
disk-like flow, but it is not clear that their SPH
calculation had the resolution to see the KH instability
and others to be discussed below. The KH instability
will also accompany RT and RM instabilities. There may
also be a coupling between the shear flow along the
top and bottom faces of the expanding torus
and the RT and RM instabilities that will be triggered
along the outer rim of the torus. It may be that the  
KH instability is mostly responsible for entrainment, 
not the growth of spokes.

The equatorial, toroidal flow will also be subject to 
RT instabilities as the denser torus
is decelerated by the less dense mantle and envelope
of the star.  The core is roughly an n~=~3 polytrope. 
The torus expands with density decreasing roughly as 
$r^{-2}$ if there is little increase in the vertical height. 
If the density profile in the core is steeper than $r^{-2}$,
the torus will always tend to be denser than the mantle
into which it expands.  While the details will depend on the 
nature (density, energy, opening angle) of the axial jets, 
the calculation of \citet{kho99} shows that the toroidal 
structure remains denser than the surrounding He core by 
about a factor of 3 as the torus comes to the edge of the 
helium core. The torus will be denser than any surrounding 
hydrogen envelope of essentially constant density. The torus 
will thus constantly be subject to RT instabilities as it 
propagates outward, depending on its heat content and
capacity to do $PdV$ work, just as for the spherical expansion 
that has traditionally been studied in this context 
\citep{chev76,kifon03,kifon06}.

While in the linear limit, small scales will grow faster by RT 
instability. In the non-linear limit the velocity of the growth 
of RT instabilities is given approximately by: 
\begin{equation}
\label{RTgrow}
v_{RT} \sim (a L)^{1/2},
\end{equation}
where {\it a} is the effective acceleration and {\it L} is the
characteristic scale length \citep{youngs86}. At this
stage, larger scales will grow faster. Laser-induced 
production of 3D RT instabilities has shown that RT 
fingers tend to form and grow at nearly the effective 
``free-fall" rate, the velocity of the interface before
deceleration, in this case $v_t$, the velocity of
the expanding torus  \citep{drake04}.

The characteristic length scale for the RT instability
is often taken to be the pressure scale height in the 
star \citep{chev76}. Alternatively,  
the characteristic length scale for the RT structures may 
be the effective vertical height of the expanding torus. 
If so, the torus might tend to break up into $\sim r/h$ 
fingers, where $r$ is the radius of the torus and $h$ its 
(full) thickness.  The jet-induced models of \citet{kho99} 
give an opening angle of $\sim$ 30\degree\ as the torus
impinges on the outer helium core. This would give 
$r/h \sim  5 -6,$ implying 5 or 6 ``fingers" per hemisphere
or 10 - 12 around the full perimeter of the torus.   This
is somewhat larger than, but perhaps in the ball park of, 
the number of ``fingers" seen in Cas A. The most prominent 
fingers might be fewer than the total. 

Equation \ref{RTgrow} depends on the amplitude of the deceleration,
which is itself a function of the density of the torus
and the ambient medium into which it propagates. Invoking
momentum conservation (neglecting the pressure within the
torus as if it were already cold and expanding homologously)
and expressing the effective acceleration as the ram pressure 
exerted by the mantle, $ \sim \rho_{man} v_t^{2}$ divided by the 
mass per unit area of the torus that is decelerated in a given
time, one can write:
\begin{equation}
\label{RTaccel}
a \sim \frac{\rho_{man} v_t^{2}}{\rho_t v_t \tau} \sim \frac{\rho_{man} v_t}{\rho_t \tau},
\end{equation}
where $\rho_{man}$ is the density of the mantle, $\rho_t$ is the 
density of the torus, $v_t$ is the velocity of the torus,
and $\tau$ is the timescale of the growth of the RT structure.
From Eqns. \ref{RTgrow} and \ref{RTaccel}, we can write 
\begin{equation}
\tau \sim \frac{\rho_{man} v_t}{\rho_t a} \sim \left(\frac{L}{a}\right)^{1/2}, 
\end{equation}
and hence
\begin{equation}
a \sim \left(\frac{\rho_{man}}{\rho_t}\right)^{2}\frac{v_t^{2}}{L},
\end{equation}
giving
\begin{equation}
v_{RT} \sim \frac{\rho_{man}}{\rho_t} v_t,
\end{equation}
where $L \sim r_t/5$ and $r_t$ is the radius of the torus. 

For the particular simulation of the core of a 15 \m\ star of 
\citet{kho99}, the torus hits the edge of the helium core of 
radius $2\times10^{10}$ cm in 35 seconds and hence has a mean 
velocity of $\sim 6000$ \kms.  The torus has a density about 
3 times that of the mantle so the acceleration that drives 
the RT instability at that point will be $a \sim 10^7$ cm s$^{-2}$
and the growth rate of the fingers will be $v_{RT} \sim \frac{1}{3}v_t$.
The latter result suggests that the RT fingers might not have
reached the saturation limit at this epoch. Note that if 
the fingers grow at the rate of motion of the interface, $v_t$
\citep{drake04}, then they could develop $\sim$ 3 times faster.    

The RM instability occurs when shocks propagate 
down density gradients and especially when shocks encounter 
density discontinuities, such as found at composition boundaries 
in massive stellar cores. For the cores of stars that were
the likely progenitor of Cas A, these are the Fe/Si boundary,
the Si/O boundary, the C-O/He boundary, and the He/H boundary,
if any \citep{kifon03,kifon06}. For perturbations of initially small 
amplitude, $\delta r$, of characteristic length, {\it L}, the growth 
rate for RM instabilities is:
\begin{equation}
v_{RM} = 2 \pi v_t \left(\frac{\delta r}{L}\right) A,
\end{equation}   
where {\it A} is the Atwood number, 
$A = \frac{\rho_t - \rho_{man}}{\rho_t + \rho_{man}}$ \citep{youngs86}.
At the boundary of the helium core, where $A \sim$ 1/2 then
$v_{RM} \sim 3 v_t \sim 9 v_{RT}$ for $\delta r \sim  L$. 

To illustrate the capacity of a jet-induced model to produce this
variety of instabilities (if not literally the structure of Cas A),
we show in Figure \ref{jets} the result of a two-dimensional simulation
of equal and opposite jets propagating from the center of a
helium star. The progenitor model was the 3.5 \m\ core remaining
from the evolution of a star of 10 \m\ \citep{woos95}.  
The jets were introduced as an inflow boundary condition at the inner
radial coordinate of 3.82$\times 10^8$ cm with an opening angle of
about 25\degree.  The jet density, pressure and injection velocity rise  
linearly to maximum values of 6.5$\times 10^5$ g cm$^{-3}$, $10^{23}$  
ergs cm$^{-3}$ and 3.22$\times 10^9$ cm s$^{-1}$, respectively, over  
the course of 0.05 seconds.  After about 0.5 seconds, the jet  
injection velocity is linearly reduced to zero in about 1.5 seconds.  
These parameters were chosen so as to deliver a  
total energy of $10^{51}$ ergs to the star (one-half that energy in
each jet). The mass interior to the lower boundary, 1.593\m, was 
included as a point mass source of gravitation. The self-gravity 
of the remaining portion of the star was neglected. These conditions 
were intended to approximately reproduce those of \citet{kho99}.

The resulting dynamics were computed using the FLASH code \citep{fryx00}. 
We used a grid in spherical coordinates with eight levels of refinement 
in both radius and angle. The maximum effective number of angular zones 
is 1024 and the maximum number of radial zones is 8192. This gives a 
minimum resolution in radius of $\sim 1.5\times10^7$ cm and in angle of 
0.175\degree (corresponding to $3\times10^6$ cm at R = $1\times10^9$ 
cm). This compares to the simulation of \citet{kho99} that had a 
minimum resolution of $\sim 3.7\times10^7$ cm near the inner boundary 
that degraded in the Cartesian grid of that simulation to
$\sim 2.3\times10^9$ cm at the outer boundary ($\sim 1.5\times10^{11}$). 
Our maximum resolution remains constant in radius and our angular 
resolution is higher throughout the simulation compared to that of 
\citet{kho99}. The instabilities that set in at larger radius are 
thus better resolved. 

Figure \ref{jets} shows the results after 47 seconds with the jets
in the vertical directions. The leading shock induced by the jets 
has propagated out of the star (initial radius $1.07\times10^{11}$ cm) 
and off the grid ($1.2\times10^{11}$ cm). The effect of the jets
in the inner core have ``healed" due to transverse pressure gradients. 
The remnants of the passage of the jets can still be seen in the 
outer triangular regions at a radius of $4 - 7\times10^9$ cm.  
The effects of instabilities are plainly seen. The RM instabilities 
and associated KH instabilities induced as the jet shock propagated 
across the Si/O boundary are seen in the ``cap" at about
$9\times10^{10}$ cm. The horizontal structure is in the equatorial 
plane and results from the convergence of the jet-induced blast 
waves on the equator. KH rippling is seen at about $5\times10^{10}$ cm 
and an RM/KH ``mushroom" is seen on the leading edge of the toroidal 
structure at about $8\times10^{10}$ cm.  All this structure would 
be different for 3D (the small extensions right on the computational
axis at $9\times10^{10}$ cm are surely artificial) and with a true, 
rather than numerically generated, Reynold's number in the flow, 
but the basic presence of the instabilities is well illustrated. 
We note that while a plausible, but arbitrary, perturbation must be 
added to induce RM and RT growth in a spherical model \citep{kifon06}, 
a jet-induced model, by assumption, represents a large asymmetric 
``perturbation," leading to ``natural" and significant instabilities. 
Details of this simulation will be presented elsewhere.
 
The RT and RM (and KH for that matter) instabilities tend to lead to 
mixing in the non-linear limit. If we want to explain ``fingers" in 
Cas A, then it may be necessary for the fingers to break out of the 
star before that turbulent, non-linear limit is reached, although
finer scale structure might be related to the fast-moving knots. The 
analysis given above (Eqns 8 - 11) suggests that might be the 
case for the RT instability; that the RT fingers have not reached 
saturation before the explosion leaves the helium core. The RM 
instabilities may grow faster locally at the composition discontinuities, 
but it is not clear in the absence of an appropriate simulation how 
this will interact with the RT structure. Limitation of the non-linear 
turbulence might be aided by the fact that the progenitor star had 
lost most or all of its hydrogen envelope and hence was something 
similar to a bare helium core. Figure \ref{jets}, which is not yet 
in homologous expansion, hints that some of this structure may survive.

We note that in the proposed picture the toroidal flow 
and ``fingers" should be composed of elements such as 
Si, Ca, O, and should be relatively devoid of iron-peak 
matter. This might agree with the high-velocity oxygen-rich 
knot structure of \citet{fesen06c} and the relative
paucity of Fe in the direction of the NE ``jet" observed
in the X-ray. Once again, we note that especially fast
moving matter in the ``fingers" may undergo adiabatic
expansion and cooling so that the bulk of it may
be hard to detect.

If there are equatorial ``fingers" of faster flow, this
flow may also drive lateral pressure waves piling
up matter on the ``walls." These walls might thus be 
denser and that higher density might inhibit the 
flow adjacent to the primary flow in the finger.
This might account for the marked gaps in the flow 
pattern reported by \citet{fesen06c}.

Another way to induce RT instabilties is to 
accelerate denser material by lighter material.
This might occur if there is a later injection of
energy from the new-born neutron star into the
previously exploded and expanding material. There
are a variety of mechanisms that might provide
a somewhat delayed input of energy into increasingly
less dense matter surrounding a new-born neutron 
star. 

\citet{wa07} have pointed out that non-axisymmetric
instabilities in rapidly rotating neutron stars
are likely to generate a strong magnetoacoustic flux.
This flux would push on the inside of the expanding
torus produced by a previous phase of MHD jets 
that propagated up the rotation axis. The resulting
interaction  might produce RT instabilities. \citet{wa07} 
note that this process is likely to be amplified by the
deleptonization, contraction and spin-up of the
proto-neutron star on the timescale on which
the radius contracts, of order 0.1 - 1 s. It is
not clear how any such magnetoacoustic flux
would be propagated: some might go up the rotation
axis, but a substantial portion might go out
the equatorial plane, directly accelerating
the postulated expanding torus from within.     

At somewhat later phases, after the neutron star has
shrunk to its final radius, $\sim10$ km, it will
continue to cool by neutrino emission for another
10 - 100s, the (other) Kelvin-Helmholtz timescale.
During this phase the neutron star may emit a 
neutrino-driven wind \citep{qian96}. Toward
the end of this phase, as the density declines in the 
vicinity of the neutron star, the flow may become 
magnetically-dominated, thus resulting in a relativistic 
wind flow \citep{bucc06,bucc07} \citep[see also][]{tho04,metz07}.  
\citet{bucc07} note that this sort of flow seems to
naturally occur in the equatorial plane. This work does 
not account for the fact that the supernova ejecta will 
not have dispersed on the relevant timescales and thus 
will present a different outer boundary condition - a 
working surface - than that assumed in the numerical models
as discussed by \citet{kom07}. Nevertheless, there is a suggestion 
for a natural tendency for an original MHD jet flow up the axis 
to evolve to an equatorial flow at later times driven
by a pulsar wind. The characteristic timescale, 10 - 100 s 
is roughly the time it takes for the original shock, 10 s, and 
the torus, somewhat slower, to propagate to the edge
of a helium core. Thus, at about the time this jet/torus
structure breaks out of the core, there might be a 
strong, relativistic, equatorial pulsar wind that
begins to accelerate the jet-induced torus from within.
The low densities associated with this late-time
fast wind will naturally lead to RT instabilities
as the slower, denser torus is accelerated from within 
by the pulsar wind. We argue that such a wind should be 
expelled into structure roughly similar to Figure \ref{jets}.

In effectively free-streaming calculations, \citet{bucc07} find
that their numerical models give essentially the same asymptotic
solutions as analytic solutions with the energy scaling like
$\sin^2 \theta$. This would mean that half the energy is
directed within a half-angle of $\sim 50$\degree. This energy
distribution is wider than the torii computed here and by 
\citet{kho99}, but an appropriate calculation must consider 
the interior of the supernova as the appropriate ``outer" 
boundary condition. This might lead to a more confined equatorial 
flow. \citet{bucc07} also note a tendency to form ``plasmoids" 
with characteristic angles of 5 - 10\degree. Such structures might 
see RT instabilities and give somewhere of order 20 - 40 ``fingers." 
The calculations of \citet{bucc07} were done for a magnetar-like 
case and the results for the density and Lorentz factor of the 
pulsar wind will depend on the specific magnetization and inner 
boundary condition, but the principles should apply to both 
pulsars and magnetars.  We note that arguments have been made 
that the compact object in Cas A is a magnetar, but also that 
there is no current obvious evidence for a pulsar wind
\citep{fesen06a,krause05}.

\section{Discussion and Conclusions}

Despite attempts to deconvolve the three-dimensional structure, 
the interpretation of the kinematics, dynamics, and morphology of 
Cas A remains difficult. We adopt a jet-induced structure
as an organizing principle to frame the analysis of observations
and to provide a model to test. We suggest that Cas A exploded from 
a jet-like mechanism, but that, unlike many interpretations, the 
main jet is not the well-known NE ``jet." We argue that a structure 
to the SE with a position angle of $\sim 125$\degree\ represents
the main jet. This feature is less immediately discernable because 
of projection effects, but is suggested in Doppler maps. It
is also iron rich. In the jet-induced model, the bulk of the
ejecta, the intermediate-mass elements, should be ejected in
an equatorial torus (for jets of equal and opposite momentum)
that results from the convergence of blast waves on the equator. 
We interpret the NE ``jet" and other structure as flows that are 
roughly normal to the main SE jet that result from non-axisymmetric 
structure in the equatorial plane of the jet. This structure may resolve 
one of the current mysteries of Cas A, that the proper motion of 
the compact object seems to be nearly normal to the principle axis 
of the dynamics. In our proposed morphology, the compact object has 
been ``kicked" at an angle $\sim 30$\degree\ to that of the main 
jet. The proposed picture also opens the way for a new interpretation 
of chemical structure, especially the fact that iron is seen beyond 
silicon: this observation may be significantly affected by projection 
effects, rather than resulting from true ``mixing" of the ejecta.   

There have been hints of the geometry we propose here in 
previous discussions of Cas A. \citet{markert83} proposed an inclined
ring model (their Figure 5), but their solution was a ring with its
axis virtually along the line of sight. To account for the SE feature, 
they invoked non-uniform emissivity around the perimeter of the ring.
The ring may have non-uniform emissivity, but their particular 
solution does not account for why this SE feature is iron-rich, 
and does not seem to be quantitatively consistent with subsequent work. 
\citet{willingale02} used {\it XMM Newton} composition and Doppler data to 
do a 3D deconvolution, presenting a proposed side view from the East 
(their Figure 11). They remark that the data is well characterized by 
the doughnut shape suggested by \citet{markert83}, but their side view 
seems to show the axis of the torus tilted at about 45\degree\ to the 
line of sight, rather than the 87\degree\ advocated by \citet{markert83}. 
With the spatial resolution available, the projection of 
\citet{willingale02} shows no particular evidence of the NE jet. 
\citet{willingale03} propose a clumpy toroidal geometry in which the 
axis of the toroid is oriented 40\degree\ East of the line of sight
and 125\degree\ (counterclockwise) to the South of North, essentially 
the same as we have chosen here. Along this torus they identify 
concentrations of mass and energy on opposing sides to the ``North" 
and ``South" that are near to, but not quite the same as North and 
South on the plane of the sky (see their Figure 4 for the coordinate 
system they define). They refer to these concentrations of mass and 
energy as ``jets" but do not relate them to the Doppler feature in 
the SE discussed by \citet{willingale02}, and they are distinctly not 
the directions of the classic NE and SW jets. It is also not quite 
clear how this torus relates to the deconvolution given by 
\citet{willingale02}. \citet{willingale03} suggest that the torus 
they identify may be the shell of matter ejected by a progenitor 
Wolf-Rayet star. Here we raise the possibility that the torus 
is that produced by a jet-induced explosion.  The mass map presented 
by \citet{willingale03} does not show any special evidence for the 
iron-rich mass in the SE that we identify as the main direction of 
the jet axis. This may be related to the fact that the hot iron 
moving along the jet axis has relatively little mass.  Like 
\citet{willingale03}, we have no immediate explanation for the 
asymmetric distribution of mass around the perimeter of this torus, 
but propose that this is related to the instabilities and 
non-axisymmetric flows normal to the jet axis that we have discussed 
here, rather than ``jets" {\it per se}. \citet{willingale03} point out 
that the apparent asymmetry of the explosion might suggest shear 
and hence turbulence and clumpiness intrinsic to the ejecta without 
needing a collision with an external medium. We have illustrated 
just this sort of effect in Figure \ref{jets}.

\citet{krause05} have provided another interesting and mysterious
aspect concerning the nature of Cas A. They have identified an
apparent bi-polar flow and associated infrared echoes. Their
preferred orientation is a position angle of about 26\degree\
at an angle of about 82\degree\ to the line of sight, nearly
in the plane of the sky. This angle differs from every other
geometric feature discussed in this paper. \citet{krause05} 
suggest that Cas A contains a magnetar that had a soft
gamma-ray repeater outburst circa 1952. There is no sign
of a Crab-like synchrotron nebula, which is also an issue
for our suggestion that a pulsar wind might play a role in
shaping the ejecta. The energy injected by a soft gamma-ray
repeater burst is rather small compared to a supernova kinetic energy, 
but such a burst of energy might affect portions of the ejecta, 
enhancing previous irregularities. It would be interesting to 
look for other possible indications of such a magnetar burst.
We note that the ionization ages identified by \citet{willingale03}
are on the scale of a century or less. 
   
\citet{burrows05} proposed that the main axis of the explosion of
Cas A is that associated with the ``iron-rich, mass-rich, 
energy-rich" SE direction and that this direction be a rotation axis. 
They speculate that the rotation axis precessed from the generation 
of the main jet to the generation of the later, secondary flow 
associated with the NE jet, or that the latter could be
a magnetic dipole axis (although a magnetic axis would be expected
to rotate in direction at the pulsar period if tilted with respect 
to the rotation axis). \citet{burrows05} argue that the inferred 
direction of motion of the central object is along the proposed 
axis of explosion with an inferred blue-shift. We have made the
specific proposal of a displacment of
the line of motion of the compact object by $\sim$30\degree\
from the axis of the main jet. \citet{burrows05} suggest
that the motion of the compact object could be due to a
slightly top-bottom asymmetric bi-polar flow. \citet{janka05}
also suggest that the NE jet is induced after the main 
explosion, perhaps due to accretion onto the neutron star.
Here we propose that there is an explicit jet/torus structure
of to Cas A, and agree with the general suggestions of \citet{burrows05}
and \citet{janka05} that the NE jet is some form of secondary
flow, caused by the complex interaction of the jets or by the
late-time effect of a pulsar wind, or perhaps repeated soft
gamma-ray repeater outbursts.    

To produce a robust explosion, the main explosion engine should 
involve $\sim 10^{51}$ ergs. Even if this energy begins in jets,
it is spread through a large solid angle by subsequent
dynamical interaction with the surrounding progenitor star. 
Thus it is not clear how much energy should be directed along
the main axis once free expansion is attained. This can be
quantified by suitable models. It remains to be seen whether 
an analyis of the SE structure similar to that done by 
\citet{laming06} will reveal a sufficiently large kinetic energy 
in the SE to be consistent with a jet in that direction being the
primary origin of the explosion. The evidence that only $\sim 10^{50}$ 
ergs is involved in the NE jet is consistent with that being only 
a secondary effect.  It is also important to determine the extent 
to which projection of an iron-rich SE jet on a silicon-rich 
equatorial flow can quantitatively account for the observed 
apparent composition inversion.  

An important aspect of this model is that the non-axisymmetric
structure arises from instabilities in or near the equatorial 
plane. We have sketched possible instabilities, including
the effects of a pulsar wind. Three-dimensional hydrodynamics
calculations or even pehaps MHD calculations \citep{stone07} are 
required to determine whether or not the structure of Cas A can 
arise in a natural and physically self-consistent way from a 
jet-induced model. Such a model would have to yield a plausible 
explanation of transverse velocities up to 14,000 \kms\ in the 
direction of the NE jet \citep{fesen06c} in contrast to 
transverse velocities of $\sim$ 8,000 \kms\ along the SE 
structure (the radial velocities in both of these directions 
are more modest; \citet{willingale02}). We note that this high 
velocity in the NE direction may be an indication that this is a 
secondary flow, perhaps driven by a pulsar wind, invoking once 
again the notion that the original jet energy must be dissipated
throughout the ejecta to cause the explosion. A separate
injection of energy may be necessary to blow holes in
the original ejecta, accelerate some regions to high
velocity and produce fast moving knots. \citet{fesen06c}
note that the NE jet is like a ``spray" of ejecta knots
rather than a narrow jet. That might be an important clue 
to the underlying process. 

There is no model today that accounts for the nature of Cas A.
The positive aspects of the model we present here are that
it roughly accounts for the iron-rich SE structure, the
extension of iron beyond silicon in that direction, and
the motion of the compact object in nearly that direction. 
On the negative side, there is much structure in Cas A that
remains unexplained. Our model gives no natural explanation for 
why the NE jet and SE counter-jet seem so nicely and oppositely 
aligned, a problem for any model in which these are secondary
flows. It may also be the case that the structure seen in 
Cas A might be generated in the absence of a jet-induced
origin. This has certainly not been demonstrated. It may also
be true that the NE jet defines a symmetry axis and the compact
object has been kicked at some oblique angle due to equatorial
instabilities, but such a theory would also require elaboration. 

An important aspect going forward is to compare models with
observations. For our model we predict: a) there is enough 
mass and kinetic energy in the SE and NW directions to account 
for the explosion (some of this matter may be fast, cool and 
difficult to directly observe; much of the originally directed
energy has spread throughout the ejecta); b) there is high-velocity 
iron extending to large distances in the SE (in front) and NW
(in back); c) the 3D structure and kinematics should show an expanding 
toroidal structure, roughly normal to the SE direction we have defined 
here, consistent with the energy deposited in the progenitor
star by the jet-induced blast waves; d) the intermediate mass 
elements are concentrated in this torus; e) the apparent ``overturn" 
that places iron beyond silicon in the plane of the observer is 
largely due to differentiated, composition-dependent, bulk flow, not 
mixing due to turbulence; f) the compact object is moving toward us; 
g) the ``holes" in the remnant caused by secondary flow, including 
the NE and SW jets, are the locus of faster moving material; h) the 
walls defining the holes are in transverse motion; the holes should be 
getting larger.  We urge those analyzing the data on Cas A to put 
these predictions to the test in order to more deeply understand 
this remarkable explosion. The structure to the SE (and NW) and the 
torus defined by \citet{willingale03}, in particular, warrant much more 
detailed study. We need to better understand the kinematic and 
dynamical relation between the SE feature, the NE/SW ``jets," and 
this torus and to know whether the torus is more consistent with a 
pre-existing shell or a structure created by the supernova.

\acknowledgments

We thank Rob Fesen, Una Hwang, Martin Laming, Tracey Delaney, 
Jeff Hester, Elaine Oran, and Milos Milosavljevic for helpful 
discussions. The software used in this work was in part developed 
by the DOE-supported ASC/Alliance Center for Astrophysical 
Thermonuclear Flashes at the University of Chicago.  This work 
was supported in part by NASA Grant NNG04GL00G and NSF Grants 
AST-0406740 and AST-0707769.

\clearpage

\begin{figure}
\centering
\includegraphics[totalheight=0.3\textheight]{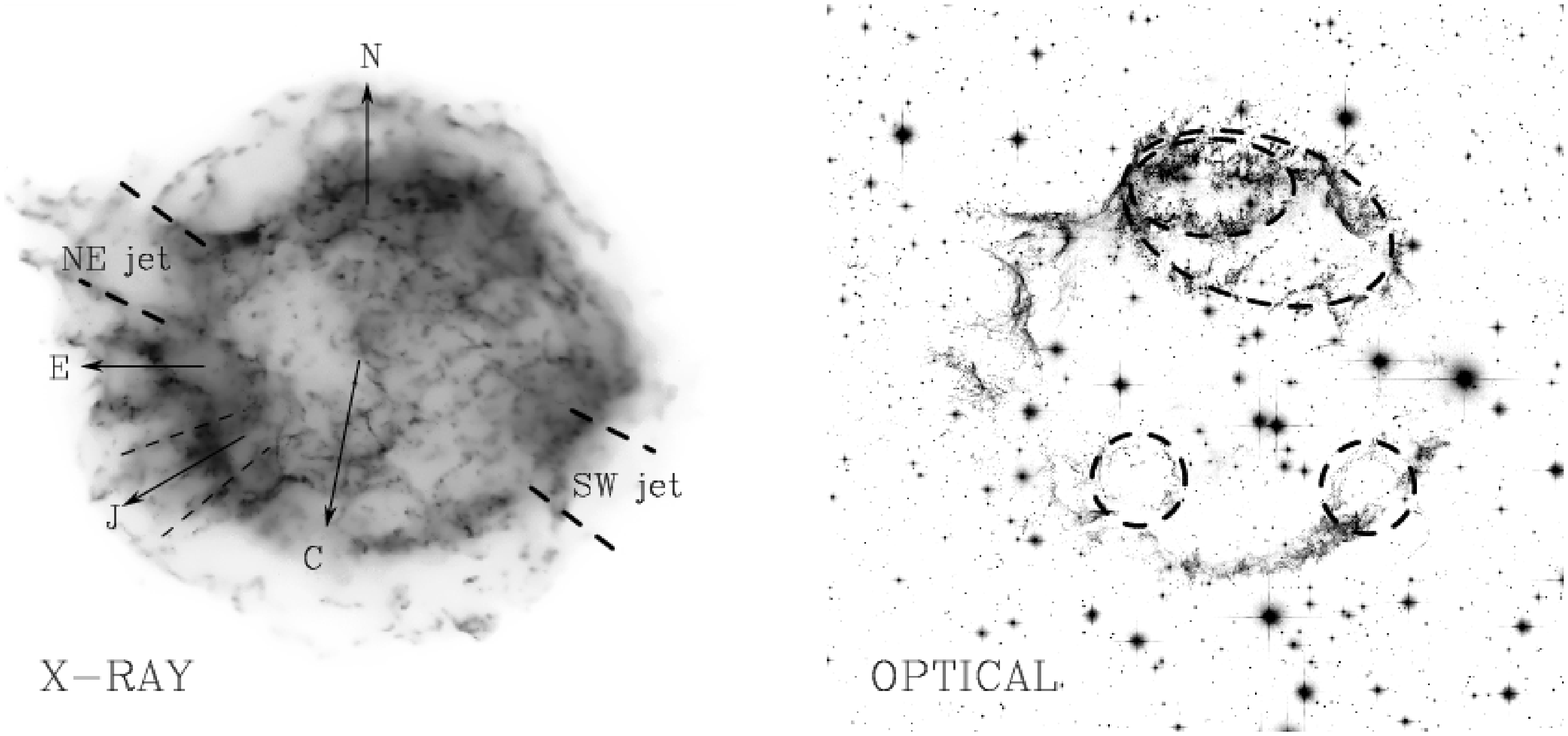}
\caption[image.eps]{(a - left) X-ray image of Cas A and 
illustration of the coordinate system for this work. The solid 
lines represent the observed position angle of the compact object, 
$\theta_{p,C}$, and that of the proposed main jet, $\theta_{p,J}$, 
with dashed lines indicating a plausible range of angles for the jet. 
Heavy dashed lines illustrate the approximate location of the NE 
and SW ``jet" and ``counter-jet" structure that we interpret here 
as a secondary flow resulting from instabilities. Adapted from 
Hwang et al. (2004). (b - right) Optical image of Cas A illustrating 
some of the ``holes" that characterize the morphology of Cas A. 
Adapted from http://heritage.stsci.edu/2006/30/index.html.   
\label{image}
}
\end{figure}

\begin{figure}
\centering
\includegraphics[angle=270,totalheight=0.6\textheight]{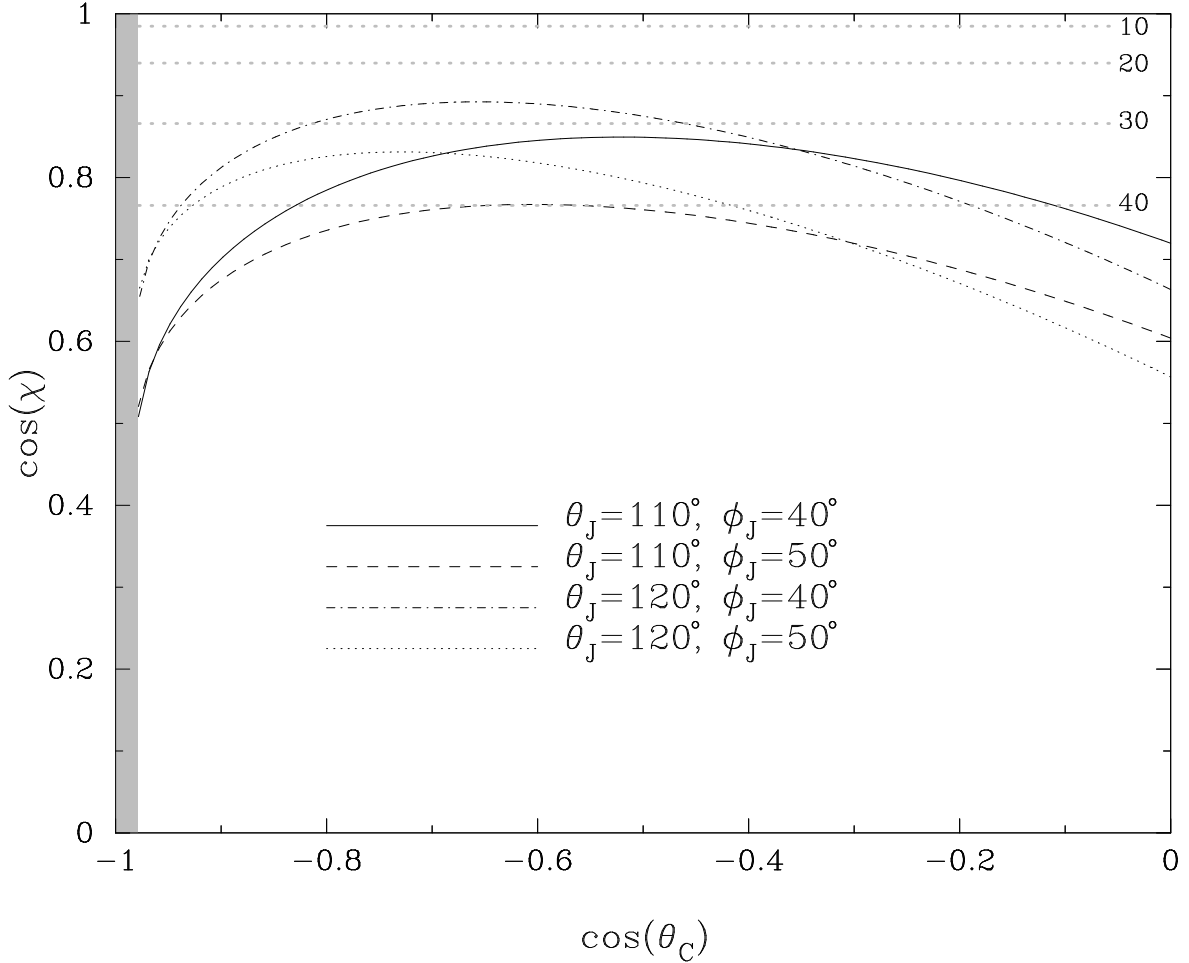}
\caption[chi.eps]{The angle, $\chi$, between the axis of the 
proposed main jet and that of the motion of the compact object 
is given as a function of the direction of recoil of the
compact object, $\theta_{C}$, for plausible choices of the 
orientation of the jet axis, $\theta_{J}$ and $\phi_{J}$.
Angles $\theta$ are measured East (counterclockwise) from North 
and angles $\phi$ are measured East from the observer line of sight
(clockwise as viewed from the North).  Horizontal lines mark 
given values of the separation angle, $\chi = 10, 20, 30$ and 
40\degree. The range of $\cos \theta_{C}$ is restricted to fall 
between 0 and -1, since the compact object is known to be 
recoiling toward the SE and that is also the assumed 
direction of the main jet in this work.  The angle $\theta_{C}$ 
is bounded on one side by $\theta_{C}$ = 90\degree\ for which 
$\phi_{C}$ = 0 and $\cos \chi = \sin \theta_{J} \cos \phi_{J}$ 
and on the other extreme by $\phi_{C}$ = 90\degree\ for which 
$\theta_{C} = \theta_{p,C}$ and $\cos \chi = \sin \theta_{J} 
\sin \phi_{J} \sin \theta_{p,C} + \cos \theta_{J} 
\cos \theta_{p,C}$.   
\label{chi}
}
\end{figure}

\begin{figure}
\centering
\includegraphics[angle=270,totalheight=0.6\textheight]{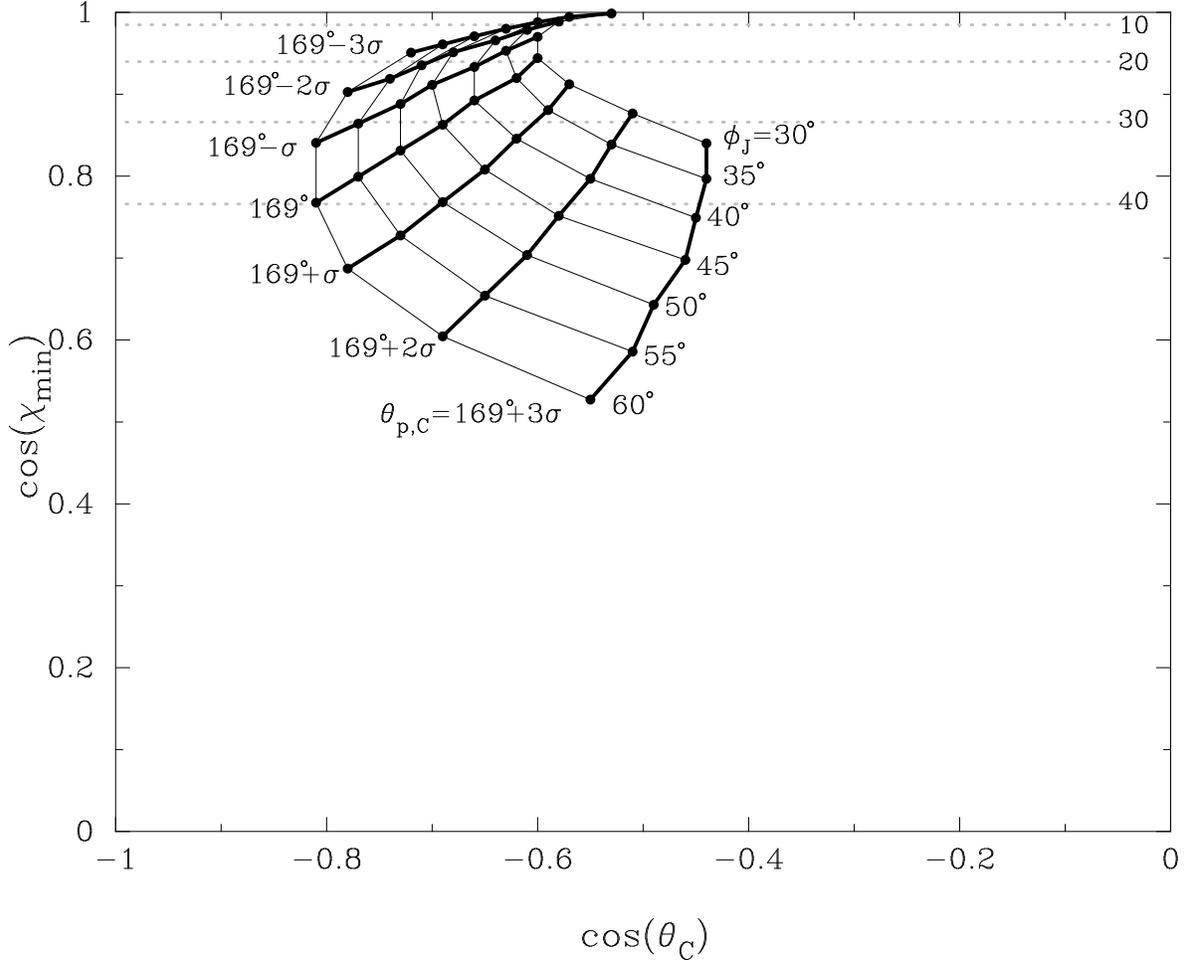}
\caption[chimin.eps]{The minimum value of the separation 
angle, $\chi_{min}$, between the axis of the proposed main jet 
and that of the motion of the compact object as shown in 
Fig. \ref{chi} is given for the orientation of the jet axis 
$\theta_{J} = 120$\degree\ and a range of plausible values 
of $\phi_{J}$, and for the nominal observed value of the 
position angle of the compact object, $\theta_{p,C} = 169$\degree, 
and 1, 2, and 3 $\sigma$ limits on that angle where
$\sigma$ = 8.4\degree. Angles $\theta$ 
are measured from the North and angles $\phi$ are measured 
from the observer line of sight. Horizontal lines mark given 
values of the separation angle, $\chi = 10, 20, 30$ and 40\degree. 
\label{chimin}
}
\end{figure}

\begin{figure}
\centering
\plottwo{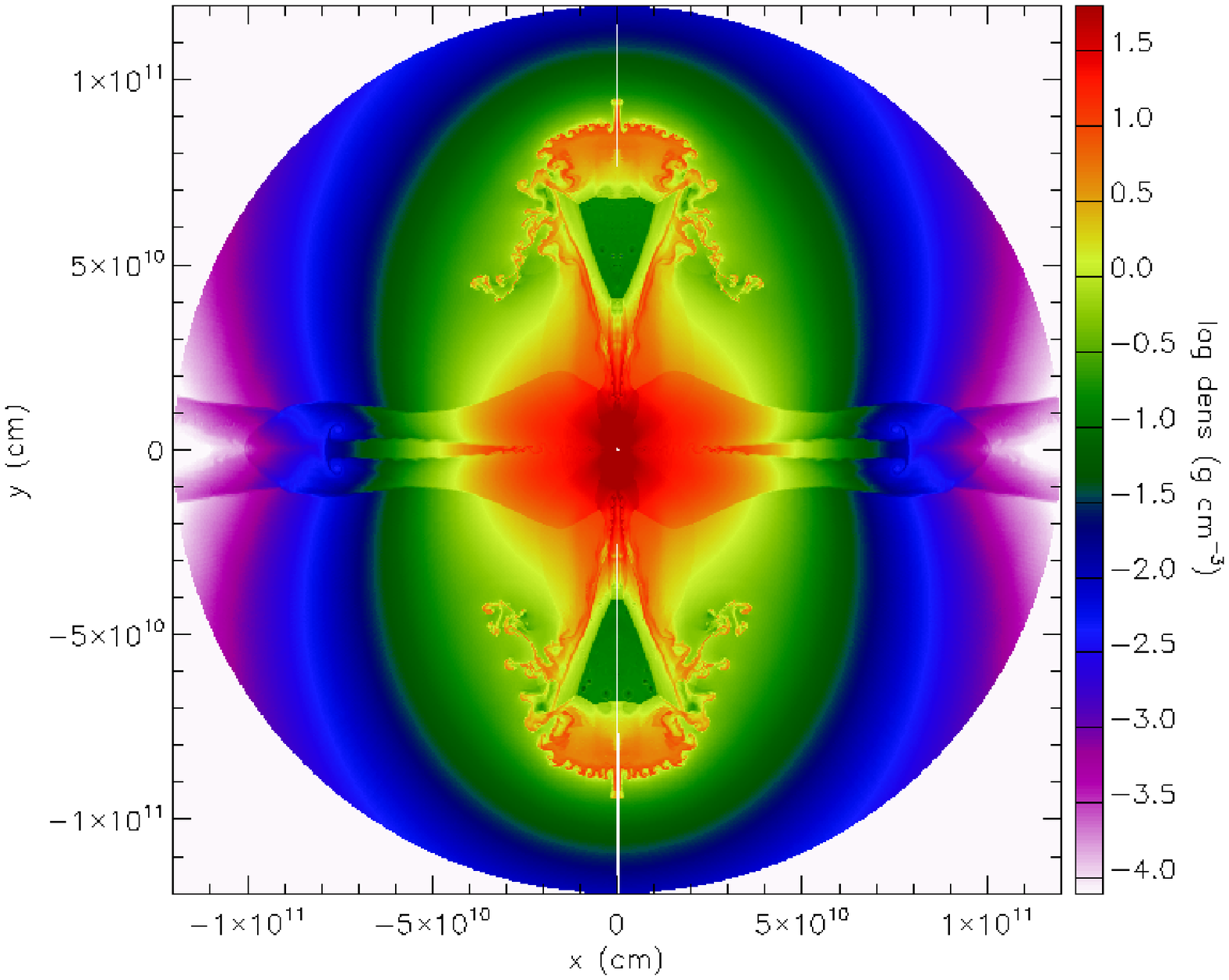}{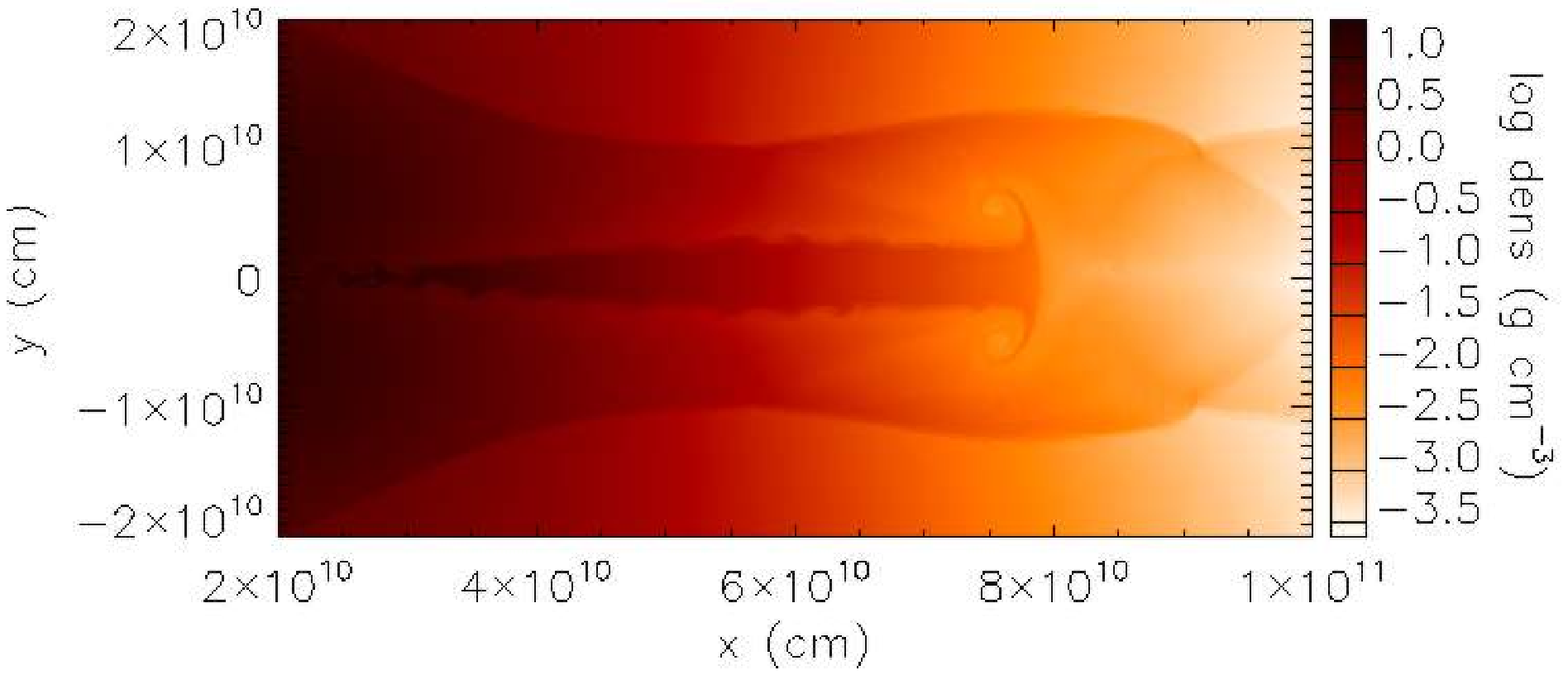} 
\caption[jet_helium]{(a - left) The result of launching symmetric vertical
jets with $10^{51}$ ergs into a helium star of 3.5 \m. This 2D 
simulation was performed with the FLASH code with a central point 
source of gravity, but neglecting the self-gravity of the remaining
matter. The computation was done on a 180\degree\ grid and a mirror
image used to create this figure. The light strip down the middle
is an artifact of this reflection process. This image corresponds 
to 47 seconds after the initiation of the explosion after the 
leading shock has left the star. The asymmetry of energy injection 
naturally leads to a variety of hydrodynamic instabilities associated 
with the direct propagation of the jets and the secondary, 
equatorial toroidal flow resulting from the collision of the 
jet-induced blast waves on the equator. (b - right) Expanded view
of the equatorial torus showing the KH ``cap" and associated
wrinkling on the top and bottom surface of the flow. See the 
electronic edition of the Journal for a color version of this 
figure.}
\label{jets}
\end{figure}


\begin{thebibliography}{}

\bibitem[Anderson et al.(1991)]{and91} Anderson, M., Rudnick, 
L., Leppik, P., Perley, R., \& Braun, R.\ 1991, \apj, 373, 146 

\bibitem[Blondin et al.(1996)]{blondin96} Blondin, J.~M., 
Lundqvist, P., \& Chevalier, R.~A.\ 1996, \apj, 472, 257 

\bibitem[Bucciantini et al.(2006)]{bucc06} Bucciantini, N., 
Thompson, T.~A., Arons, J., Quataert, E., \& Del Zanna, L.\ 2006, \mnras, 
368, 1717

\bibitem[Bucciantini et al.(2007)]{bucc07} Bucciantini, et 
al.\ 2007, ArXiv Astrophysics e-prints, arXiv:astro-ph/0701578 

\bibitem[Burrows et al.(2005)]{burrows05} Burrows, A., Walder, R.,
Ott, C. D. \& Livne, E. 2005, ASP Conference Series, 332, 350

\bibitem[Chevalier(1976)]{chev76} Chevalier, R.~A.\ 1976, 
\apj, 207, 872 

\bibitem[Chevalier \& Oishi(2003)]{chev03} Chevalier, R.~A., 
\& Oishi, J.\ 2003, \apjl, 593, L23

\bibitem[Cropper et al.(1988)]{crop88} Cropper, M., Bailey, 
J., McCowage, J., Cannon, R.~D., \& Couch, W.~J.\ 1988, \mnras, 231, 695

\bibitem[DeLaney et al.(2004)]{delaney04} DeLaney, T., Rudnick, 
L., Fesen, R.~A., Jones, T.~W., Petre, R., \& Morse, J.~A.\ 2004, \apj, 
613, 343 

\bibitem[Dewey et al.(2006)]{dewey06} Dewey, D., DeLaney, T., 
\& Lazendic, J.~S.\ 2006, ArXiv Astrophysics e-prints, 
arXiv:astro-ph/0611908 

\bibitem[Drake et al.(2004)]{drake04} Drake, R. P., Leibrandt, E. C.,
Kuranz, C. C., Blackburn, M., Robey, H. F., Remington, B. A., Edwards, M. J.
Miles, A. R., Perry, T. S., Wallace, R. J., Louis, H., Knauer, J. P.,
\& Arnett, D. 2004, Physics of Plasmas, 11, 1. 

\bibitem[Ennis et al.(2006)]{ennis06} Ennis, J.~A., Rudnick, 
., Reach, W.~T., Smith, J.~D., Rho, J., DeLaney, T., Gomez, H., \& Kozasa, 
T.\ 2006, \apj, 652, 376 

\bibitem[Fesen \& Gunderson(1996)]{fesen96} Fesen, R. A., \& Gunderson, K.
   S. 1996, \apj, 470, 967

\bibitem[Fesen(2001)]{fesen01} Fesen, R. A. 2001 \apjs, 133, 161

\bibitem[Fesen et al.(2001)]{fesenetal01} Fesen, R.~A., Morse, 
J.~A., Chevalier, R.~A., Borkowski, K.~J., Gerardy, C.~L., Lawrence, S.~S., 
\& van den Bergh, S.\ 2001, \aj, 122, 2644 

\bibitem[Fesen. Pavlov \& Sanwal(2006)]{fesen06a} Fesen, R.~A., Pavlov, 
G.~G., \& Sanwal, D.\ 2006, \apj, 636, 848

\bibitem[Fesen et al.(2006a)]{fesen06b} Fesen, R.~A., et al.\ 
2006, \apj, 636, 859 

\bibitem[Fesen et al.(2006b)]{fesen06c} Fesen, R.~A., et al.\ 
2006, \apj, 645, 283 

\bibitem[Fesen et al.(2007)]{fesen07} Fesen, R.~A., et al.\ 
2007, \apj, in press, astro-ph/0603371 

\bibitem[Fryxell et al.(2000)]{fryx00} Fryxell, B., et al.\ 
2000, \apjs, 131, 273

%\bibitem[Hammel \& Fesen(2007)]{hammel07} Hammel, X.` Y. \& Fesen, 
%R.~A. 2007, \apj, 111, 111 

\bibitem[H\"{o}flich, Khokhlov \& Wang(2001)]{hof01} H\"{o}flich, P.,
   Khokhlov, A., \& Wang, L. 2001, in Proc. of the 20th Texas
   Symposium on Relativistic Astrophysics, eds. J. C. Wheeler \& H.
   Martel (New York: AIP)

\bibitem[Hungerford et al.(2003)]{hunger03} Hungerford, A.~L., 
Fryer, C.~L., \& Warren, M.~S.\ 2003, \apj, 594, 390 

\bibitem[Hughes et al.(2000)]{hughes00} Hughes, J. P., Rakowski, C. E.,
   Burrows, D. N., \& Slane, P. O. 2000, \apjl, 528, L109

\bibitem[Hwang et al.(2000)]{hwang00} Hwang, U., Holt, S. S., \& Petre, R.
   2000, \apjl, 537, L119

\bibitem[Hwang et al.(2001)]{hwang01} Hwang, U., Szymkowiak, 
A.~E., Petre, R., \& Holt, S.~S.\ 2001, \apjl, 560, L175 

\bibitem[Hwang \& Laming(2003)]{hwang03} Hwang, U., \& Laming, 
J.~M.\ 2003, \apj, 597, 362 

\bibitem[Hwang et al.(2004)]{hwang04} Hwang, U., et al.\ 2004, 
\apjl, 615, L117 % million second

\bibitem[Janka et al.(2005)]{janka05} Janka, H.-Th., Scheck, L. 
Kifonidis, K., M\"uller, E. \& Plewa, T. 2005, ASP Conference 
Series, 332, 363

\bibitem[Keohane et al.(1996)]{keo96} Keohane, J.~W., 
Rudnick, L., \& Anderson, M.~C.\ 1996, \apj, 466, 309

\bibitem[Khokhlov et al.(1999)]{kho99} Khokhlov, A. M., H\"{o}flich,
   P., Oran, E. S., Wheeler, J. C., Wang, L., \& Chtchelkanova, A. Yu.
   1999, \apjl, 524, L107

\bibitem[Khokhlov \& H\"{o}flich(2001)]{kho01} Khokhlov, A. \&
   H\"{o}flich, P. 2001, in AIP Conf. Proc. No. 556, Explosive Phenomena in
   Astrophysical Compact Objects, eds. H.-Y, Chang, C.-H., Lee, \& M. Rho
   (New York: AIP), 301

\bibitem[Kifonidis et al.(2003)]{kifon03} Kifonidis, K., Plewa, 
T., Janka, H.-T., M\"{u}ller, E.\ 2003, \aap, 408, 621 

\bibitem[Kifonidis et al.(2006)]{kifon06} Kifonidis, K., Plewa, 
T., Scheck, L., Janka, H.-T., M\"{u}ller, E.\ 2006, \aap, 453, 661 

\bibitem[Komissarov \& Barkov(2007)]{kom07} Komissarov, S. \& Barkov, M.
2007, MNRAS, in press

\bibitem[Krause et al.(2005)]{krause05} Krause, O., et al.\ 
2005, Science, 308, 1604 

\bibitem[Laming et al.(2006)]{laming06} Laming, J.~M., Hwang, 
U., Radics, B., Lekli, G., \& Tak{\'a}cs, E.\ 2006, \apj, 644, 260 

\bibitem[Liszt \& Lucas(1999)]{liszt99} Liszt, H., \& Lucas, 
R.\ 1999, \aap, 347, 258

\bibitem[Maeda \& Nomoto(2003)]{maeda03} Maeda, K., \& Nomoto, 
K.\ 2003, \apj, 598, 1163

\bibitem[Markert et al.(1983)]{markert83} Markert, T.~H., Clark, 
G.~W., Winkler, P.~F., \& Canizares, C.~R.\ 1983, \apj, 268, 134

\bibitem[Maund et al(2007a)]{maund07a} Maund, J. R., Wheeler, J. C.,
Patat, F., Baade, D., Wang, L. \& H\"oflich, P. 2007a, \apj, accepted,
astro-ph/0709.1487

\bibitem[Maund et al(2007b)]{maund07b} Maund, J. R., Wheeler, J. C.,
Patat, F., Baade, D., Wang, L. \& H\"oflich, P. 2007b, \mnras,
accepted, astro-ph/0707.2237 

\bibitem[Maund et al(2007c)]{maund07c} Maund, J. R., Patat, F., 
Baade, D., H\"oflich, P., Wang, L. \&  Wheeler, J. C.,2007c, \aa, 
accepted, astro-ph/0709.0004

\bibitem[Metzger et al.(2007)]{metz07} Metzger, B.~D., 
Thompson, T.~A., \& Quataert, E.\ 2007, \apj, 659, 561

\bibitem[Morse et al.(2004)]{morse04} Morse, J.~A., Fesen, 
R.~A., Chevalier, R.~A., Borkowski, K.~J., Gerardy, C.~L., Lawrence, S.~S., 
\& van den Bergh, S.\ 2004, \apj, 614, 727

\bibitem[Ng \& Romani(2006)]{ng06} Ng, C.-Y., \& Romani, 
R.~W.\ 2006, \apj, 644, 445

\bibitem[Qian \& Woosley(1996)]{qian96} Qian, Y.-Z., \& 
Woosley, S.~E.\ 1996, \apj, 471, 331

\bibitem[Ryle \& Smith(1948)]{ryle48} Ryle, M., \& Smith, 
F.~G.\ 1948, \nat, 162, 462

\bibitem[Stone \& Gardiner(2007)]{stone07} Stone, J.~M., \& 
Gardiner, T.\ 2007, ArXiv e-prints, 709, arXiv:0709.045

\bibitem[Thompson et al.(2004)]{tho04} Thompson, T.~A., 
Chang, P., \& Quataert, E.\ 2004, \apj, 611, 380

\bibitem[Thorstensen et al.(2001)]{thors01} Thorstensen, J.~R., 
Fesen, R.~A., \& van den Bergh, S.\ 2001, \aj, 122, 297

\bibitem[Wang et al.(2002)]{wang02} Wang, L., et al.\ 2002, 
\apj, 579, 671

\bibitem[Wang et al.(2003)]{wan03} Wang, L., Baade, D., H\"{o}flich, P.,
   \& Wheeler, J. C. 2003, \apj, 592, 457

\bibitem[Wang \& Wheeler(2008)]{WW08} Wang, L. \& Wheeler, J. C. \ 2008, 
\araa, in press 

\bibitem[Wheeler et al.(2007)]{WMA07} Wheeler, J.~C., Maund, 
J.~R., \& Akiyama, S.\ 2007, in ``Supernova 1987A: 20 Years After,"
eds. S. Immler, K. Weiler, in press, arXiv:0704.3960

\bibitem[Wheeler \& Akiyama(2007)]{wa07} Wheeler, J.~C., \& 
Akiyama, S.\ 2007, \apj, 654, 429

\bibitem[Willingale et al.(2002)]{willingale02} Willingale, R., Bleeker,
   J. A. M., van der Heyden, K. J., Kaastra, J. S. \& Vink, J. 2002, \aap,
   381, 1039

\bibitem[Willingale et al.(2003)]{willingale03} Willingale, R., 
Bleeker, J.~A.~M., van der Heyden, K.~J., \& Kaastra, J.~S.\ 2003, \aap, 
398, 1021

\bibitem[Woosley et al.(1995)]{woos95} Woosley, S.~E., Langer, 
N., \& Weaver, T.~A.\ 1995, \apj, 448, 315

\bibitem[Young et al.(2006)]{young06} Young, P.~A., et al.\ 
2006, \apj, 640, 891

\bibitem[Youngs(1986)]{youngs86} Youngs, D.~L. 1986, Physica, 12D, 32  

\end{thebibliography}
\end{document}